\documentclass[twocolumn,english,showpacs,showkeys,reprint,linenumbers]{revtex4}
\usepackage[T1]{fontenc}
\usepackage[latin9]{inputenc}
\usepackage{textcomp}
\usepackage{amsmath}
\usepackage{graphicx}

\makeatletter
\@ifundefined{textcolor}{}
{%
 \definecolor{BLACK}{gray}{0}
 \definecolor{WHITE}{gray}{1}
 \definecolor{RED}{rgb}{1,0,0}
 \definecolor{GREEN}{rgb}{0,1,0}
 \definecolor{BLUE}{rgb}{0,0,1}
 \definecolor{CYAN}{cmyk}{1,0,0,0}
 \definecolor{MAGENTA}{cmyk}{0,1,0,0}
 \definecolor{YELLOW}{cmyk}{0,0,1,0}
}

\usepackage{babel}

\makeatother

\usepackage{babel}
\begin{document}

\title{Morphology control of the magnetization reversal mechanism in Co$_{80}$Ni$_{20}$
nanomagnets}

\author{Silvana Mercone$^{1}$}

\email{silvana.mercone@univ-paris13.fr}

\author{Fatih Zighem$^{1}$ }

\author{Brigitte Leridon $^{2,3,4}$}

\author{Audrey Gaul$^{1}$}

\author{Frédéric Schoenstein$^{1}$}

\author{Noureddine Jouini$^{1}$}

\affiliation{$^{1}$Laboratoire des Sciences des Procédés et des
Matériaux (UPR3407), CNRS-Université Paris XIII, Sorbonne Paris
Cité, Villetaneuse, France}

\affiliation{$^{2}$CNRS-UMR8213, Laboratoire de Physique et d'étude
des Matériaux, 10 Rue Vauquelin, F-75231 Paris cedex 5, France }

\affiliation{$^{3}$ESPCI-ParisTech, LPEM, 10 Rue Vauquelin, F-75231
Paris cedex 5, France }

\affiliation{$^{4}$Sorbonne Université, UPMC, LPEM, 10 Rue
Vauquelin, F-75231 Paris cedex 5, France}

\date{July 28 2014 }
\begin{abstract}
Nanowires with very different size, shape, morphology and crystal
symmetry can give rise to a wide \textit{ensemble} of magnetic
behaviors whose optimization determines their applications in
nanomagnets. We present here an experimental work on the shape and
morphological dependence of the magnetization reversal mechanism in
weakly interacting Co$_{80}$Ni$_{20}$ hexagonal-close-packed
nanowires. Non-agglomerated nanowires (with length $L$ and diameter
$d$) with a controlled shape going from quasi perfect cylinders to
diabolos, have been studied inside their polyol solution in order to
avoid any oxidation process. The coercive field $H_{C}$ was found to
follow a standard behavior and to be optimized for an aspect ratio
$\frac{L}{d}>15$. Interestingly, an unexpected behavior was observed
as function of the head morphology leading to the strange situation
where a diabolo shaped nanowire is a better nanomagnet than a
cylinder. This paradoxical behavior can be ascribed to the
growth-competition between the aspect ratio $\frac{L}{d}$ and the
head morphology ratio $\frac{d}{D}$ ($D$ being the head width). Our
experimental results clearly show the importance of the independent
parameter ($t$ = head thickness) that needs to be considered in
addition to the shape aspect ratio ($\frac{L}{d}$) in order to fully
describe the nanomagnets magnetic behavior. Micromagnetic
simulations well support the experimental results and bring
important insights for future optimization of the nanomagnets
morphology.
\end{abstract}

\maketitle

\section{Introduction}

Since the development of a wide range of synthesis routes for the
elaboration of controlled nano-objects, the number of studies trying
to understand the \textit{zoology} of the magnetic properties showed
by these systems is continuously increasing\cite{S. P. Gubin,L.
Sun,J. Zhang,K. Soulantica,Y. Soumare,G. Viau,J. Sanchez-Barriga,L.
G. Vivas,W. Fang}. Among them, Co-based anisotropic nanomaterials
are attracting a great part of scientific interest mainly for their
possible applications as nanomagnets in (nano)medicine\cite{A.
Akbarzadeh} and recording media\cite{D. Weller}, and as building
blocks in high energy nanostructured bulk magnets\cite{K. Gandha}.
All these studies showed that depending on the elaboration process,
the nanowires can present very different size, shape, morphology and
crystal symmetry as well as very different as-grown spacing between
the nano-objects\cite{J. Meier,G. C. Han,D. J. Sellmeyer,M.
Vazquez,S. J. Hurst,Q. Liu,F. Dumestre 2002,F. Dumestre 2003,H.
Zeng,J. Sanchez-Barriga}. Thus, altogether, those parameters give
rise to a wide \textit{ensemble} of magnetic behaviors that are far
from being fully controlled. For Co-based nanowires, a huge issue is
the fine control of their shape and morphology and how these
characteristics influence the magnetic properties\cite{S. P.
Gubin,L. Sun,J. Zhang,W. Fang,F. Dumestre 2002,H. Zeng,F. Ott -
JAP,F. Zighem_S. Mercone}. Indeed, the experimental probe of the
coercivity behavior as function of the shape and morphology is
mandatory for a complete understanding of the magnetization reversal
mechanism in these nano-objects and thus for their applications.

Experimental\cite{J. Zhang,J. Sanchez-Barriga,L. G. Vivas,V.
Sokalski } and theoretical\cite{C.J. Aas} proofs of the influence of
the crystal symmetry (hcp and fcc) of Co based nanostructures on the
effective anisotropy, have been widely reported for both the
magnetocrystalline term and the shape one. These terms have
demonstrated a big influence on the uniformity of the magnetization
reversal mechanism\cite{L. G. Vivas,J. Sanchez-Barriga,J. Zhang}. It
is now also well established that length and diameter size can play
an important role over the static magnetic properties of different
shaped nanowires (e.g. coercivities meltdown can be easily observed
with increasing diameter and decreasing length of the nanowires)
\cite{H. Zeng,J. Zhang,R.P. Cowburn,D. Ung,A. Gaul,F. Ott - JAP,F.
Zighem_S. Mercone}. Studies of the head-morphology effect on the
reversal mechanism are less evident and, to our knowledge, still an
experimental open issue. This is mainly due to the difficulties of
the fine control of the nanowire shape and morphology during their
elaboration. Polyol process has demonstrated in the last decades, a
very good control of the shape and crystallinity of nano-objects
based on ferromagnetic 3d-metal ions\citet{F. Fievet}. This process
has attracted a lot of attention thanks to its low cost together
with the easy and reliable control shown over the nano-objects
composition, shape and dispersion. In the case of Co based nanowires
the polyol process leads to a very specific shape and morphology
control that we used in this work in order to study into details the
shape and morphology effect over the magnetic properties of
non-agglomerated (weak interacting) Co$_{80}$Ni$_{20}$ pure
hexagonal-close-packed nanowires.

\section{Experimental details}

Non-agglomerated Co$_{80}$Ni$_{20}$ nanowires have been fabricated
\textit{via} the polyol process route which simply consists in the
reduction of metallic salts in liquid polyol\citet{Y. Soumare,D.
Ung,N. Ouar}. This method has shown an accurate size control of the
nanoparticles \textit{via} the kinetic control of the growth step
\citet{G. Viau,D. Ung}. Briefly, in this soft chemistry route, the
liquid polyol acts as a solvent and as a reducing agent for the
metallic cation. By selecting a polyol with a suitable boiling
point, the temperature reaction can be adjusted over a wide range of
temperature letting this elaboration method to be very suitable for
industrial transfer technology. In our case we used 1,2-butanediol
{[}BEG{]} polyol (boiling temperature=465K, dielectric constant=22.4
pF.m$^{-1}$). Different morphologies have been obtained by mixing
various concentration of Ruthenium chloride (RuCl3) (\textit{i.e.}
between 0.0018M and 0.0048M) to 0.15M NaOH in BEG. Once Cobalt
(0.08M) and Nickel (0.08M) precursors are dissolved in the mixture,
this latter is then heated up to 453K with a controlled temperature
ramp of 279K.min$^{-1}$ leading to Co$_{80}$Ni$_{20}$
one-dimensional nanostructures.

\begin{figure}
\includegraphics[bb=20bp 300bp 580bp 595bp,clip,width=8.5cm]{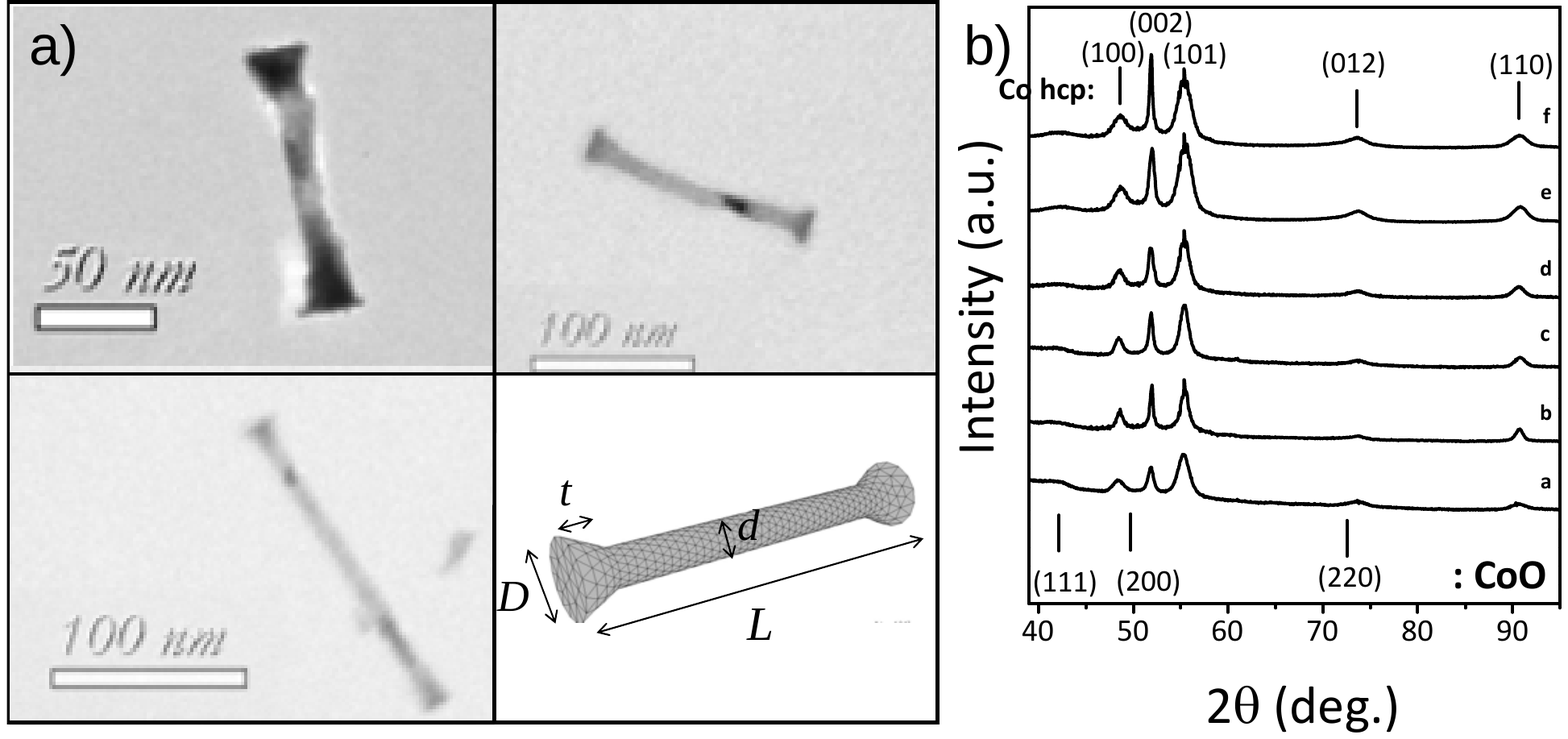}

\protect\protect\caption{(a) TEM images of various
Co$_{80}$Ni$_{20}$ nanowires elaborated under different polyol
synthesis conditions; in the right-hand bottom square is reported a
schematic mesh reproducing the typical nanowires morphology
observed; (b) X-Ray diffraction patterns for six different batches
exhibit the hexagonal symmetry (Co hcp phase is indexed). For these
measurements we had to use dry nanopowders and this is why, in some
cases, we observed typical diffraction picks of the CoO phase. }

\label{Fig_Nano-objets}
\end{figure}

Standard TEM imaging and X-Ray diffraction (XRD) patterns have been
performed for all the elaborated systems. Results obtained for some
of the elaborated nanowires are reported in Figure
\ref{Fig_Nano-objets} (a) and (b). TEM images highlighted the
necessity of four appropriate morphological parameters: i) the
diameter $d$ in the middle of the nanowire, ii) the length
\textit{$L$} of the nanowire, iii) the width \textit{$D$} of the
head of the nanowire and iv) the thickness \textit{$t$} of the head.
Accurate measurements of these four morphological parameters have
been performed by measuring more than 150 pictured nanowires per
batch. Histograms of each parameter \citet{Supplementary}(not shown
here) have shown typical log-normal distribution confirming for each
batch a good monodispersion. Typical dispertion on $D$ , $t$ and $d$
parameters has been found of $\pm 1$ nm while for $L$ the dispertion
could be larger and depending on the batch could be found up to $\pm
10$ nm. The good control of the morphology by the polyol synthesis,
allowed us to obtain very long nanowire ($L\gg d$) as well as short
ones ($L\sim2d$) (see Figure \ref{Fig_Nano-objets}(a)) with
morphologies varying from the cylinder type (\textit{i.e. }$d=D$) to
the diabolo one (\textit{i.e. }$D>d$). TEM analysis gives typical
morphological parameters laying in the following ranges:
$L=\left(65-160\right)$ nm, $D=\left(8-24\right)$ nm,
$t=\left(7-14\right)$ nm, $d=\left(7-12\right)$ nm
\citet{Supplementary}.

X-Ray diffraction patterns have shown pure hexagonal close-packed
(hcp) crystal phase for all elaborated nanowires (see Figure
\ref{Fig_Nano-objets}(b)) with presumably \textit{c}-axis of the
hexagonal symmetry (\textit{i.e.} (001)) parallel to the nanowires
length.

Using a Quantum Design MPMS 3 magnetometer, we measured standard
zero field cooled (ZFC) and field cooled (FC) hysteresis cycles at
different temperatures. Each ZFC magnetic loop was performed after
cooling down the nanowires in the polyol solution under zero
magnetic field while the FC loops were performed after cooling the
same sample under the application of $H=70$ kOe. Note that, in this
latter conditions a huge percentage of nanowires is expected to
align along the applied magnetic field before the polyol solution
freezes. Indeed all our FC cycles have been performed with a
magnetic field parallel to the length\textit{ $L$} of the blocked
nanowires (\textit{i.e.} magnetic field parallel to the easy axis of
the nanowires). The probed temperature range was between 200K and
20K. The upper temperature limit was fixed by the freezing
temperature of the 1,2-butanediol solution (the $T_{G}$ of the
{[}BEG{]} is of 220K). It is well-known that in the case of dry
powders of nanowires, a CoO nanometric shell appears on the surface
of the Co nanowires giving rise to a characteristic shift of the FC
cycles at low temperature. This shift is due to exchange bias effect
coming from the coupling of the AFM CoO or NiO shell together with
the FM Co$_{80}$Ni$_{20}$ core \citet{F.Ott_PRB}. Keeping the
nanowires in polyol solution avoids any air exposure and therefore
prevents metal from oxidation.In this condition we could verify the
absence of the exchange bias phenomenon from the magnetization
reversal inside the nano-objects \citet{F.Ott_PRB} (\textit{i.e.} no
shift of the hysteresis cycles was observed even at low temperature
\citet{Supplementary}). A straightforward correlation between the
coercive field and the shape and morphology of the nano-objects is
thus possible. Below 200 K the nanowires were (mechanically) blocked
in the freezed ``mother'' solution, allowing to exclude mechanical
reversal of the nanowires while applying an external magnetic field.
Finally, keeping the nanowires inside the polyol synthesis solution
allows to assume non-agglomerated nano-objects and equal spacing
(\textit{i.e.} isotropic interactions) between the nanowires of
different batches. This latter characteristic is very useful while
comparing static magnetization reversal mechanism among
differently-shaped nanowires.

\section{Nanowire Shape and length effects}

\begin{figure}
\includegraphics[bb=10bp 150bp 290bp 550bp,clip,width=8.5cm]{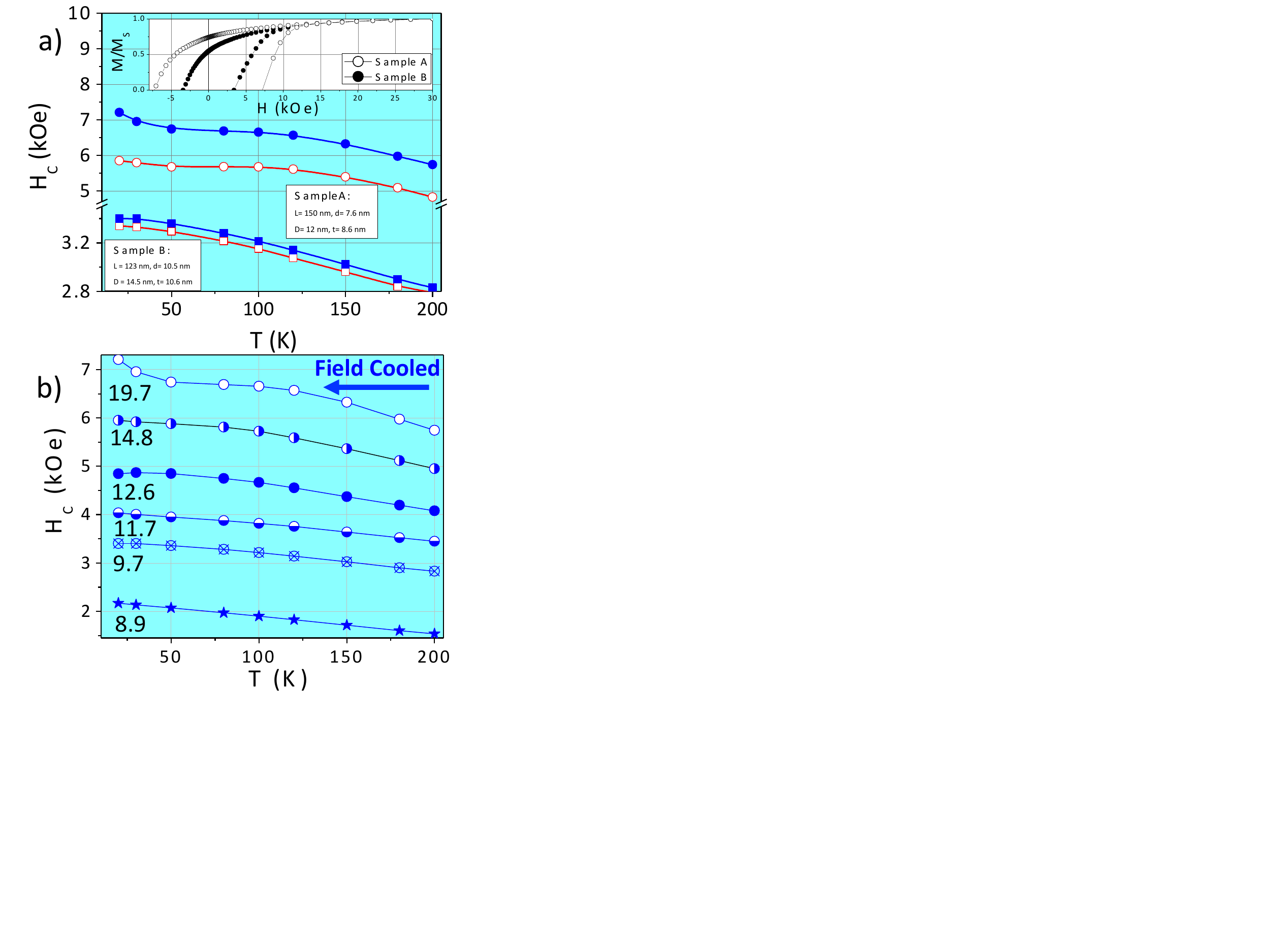}

\protect\protect\caption{a) ZFC (open symbols) and FC (filled
symbols) coercive temperature behavior for two different
morphologies (sample A cylinder type (circle symbols) and B diabolo
type (square symbols)) among all the studied nanowires (the
morphology parameters are reported for each case next to the
experimental points); upper inset shows the half-positive hysteresis
normalized cycles ($\frac{M}{M_{S}}$ ) for both A and B samples. b)
FC coercive behavior for different shaped nanowires (values reported
in the left side of the graph represent the $\frac{L}{d}$ ratio for
each coercive curve). }

\label{Fig_2}
\end{figure}

ZFC and FC hysteresis loops at a fixed temperature have been
analyzed in terms of coercivity \citet{Supplementary}). Because of
the fact that the nanowires were kept in their polyol solution,
neither the volume nor the mass of nanowires could be correctly
estimated, making it impossible to extract quantitatively the
saturation magnetization. Thus, only the coercive field analysis
will be reported thereafter. As explained in the previous section,
the studied samples presented typical cylindrical to diabolo-like
shapes with more or less pronounced head morphology (see Figure
\ref{Fig_Nano-objets}(a)). We could then study the morphology effect
on the magnetization reversal versus nanowire controlled shapes
going continuosly from the well defined diabolo (TEM images of the
upper line in Figure\ref{Fig_Nano-objets}(a)) to the cylinder type
(left-bottom-hand TEM image of Figure\ref{Fig_Nano-objets}(a)). In a
very primitive senario, the coercive field $H_{C}$ in this Co-based
one-dimensional systems is mainly driven by an effective anisotropy
field composed by two main contributions: i) one belonging to the
magnetocrystalline anisotropy ($K_{MC}$) and ii) the other related
to the shape term ($K_{Shape}$). The first term is expected to be
temperature dependent while the second is almost constant as
function of temperature (in the range 20-200 K) and only differs for
different morphologies \citet{AitAtmane2013}. The effective
anisotropy acting then on a single nanowire can be written as the
sum of the magnetocrystalline and shape contributions if both are
aligned with the length of the nanowire. Thus the coercivity can be
defined as:

\[
H_{C}\approx
H_{MC}\left(T\left(K\right)\right)+H_{Shape}\left(L,d,D,t\right)
\]
 In Figure\ref{Fig_2} (upper graph) typical ZFC and FC coercive behavior
is reported for two of the different studied morphologies. The value
of the four morphological parameters for the two samples are
reported in the graphs. The coercive behavior of both samples versus
temperature displays the variation expected for the
magnetocrystalline anisotropy contribution. Indeed the coercive
field increases with the decreasing temperature. All our nanowires
show a pure hcp structure which is usually linked to a high
magnetocrystalline term $(K_{MC}\approx5\times10^{6}$ erg.cm$^{-3}$
at room temperature), approximately linearly dependent on
temperature. Considering this term equal for all the nanowires, we
can then ascribe the observed differences in the coercive field
values only to the shape contribution.

Before going through details into this morphology differences, let
us comment on another important feature shown in the upper graph of
Figure \ref{Fig_2}, which is the ZFC behavior compared to the FC
one. As expected for non-agglomerated nanowires, the FC coercive
values are systematically higher than the ZFC ones, proving the
efficiency of the nanowires alignment by the magnetic field
application during cooling . The two samples proposed in Figure
\ref{Fig_2} (upper graph) present a substantially different
alignment effect under the FC procedure. The gap between the FC and
the ZFC coercive values for each sample is appreciable. In the case
of longer nanowires (circle symbols) this gap is relatively high
(around 1kOe) all along the temperature range while in the case of
the shorter nanowires (square symbols) the gap is much smaller
(around 100Oe). This leads us to conclude that the longer nanowires
($L=150$ nm) align better along the magnetic field than the shorter
ones. This result is pretty much coherent with a one-dimensional
nano-object having the anisotropy axis parallel to the long axis. At
this point, we should also notice that the longer nanowires present
also smaller diameter ($d=7.6$ nm ) and smaller head sizes ($D=12$
nm and $t=8.6$ nm) which means that their shape is closer to a
cylinder, while shorter nanowires are closer to a diabolo shapes.
All those characteristics are favorable for the observation of
higher coercive field values once the nanowires are aligned along
the external applied magnetic field (\textit{i.e.} FC
values)\citet{L. Sun,K. Soulantica,H. Zeng}. These results suggest
that when the shape of the nanowires approaches the cylinder type,
it is easier to align them along the magnetic field and their
coercive values go closer to the expected ones for Co based
nanowires (expected value of 8.8 kOe at room temperature). On the
contrary, when approaching the diabolo shape (square symbols in
Figure \ref{Fig_2}), the nanowires alignment seems to become less
efficient (\textit{i.e.} the gap between the ZFC and FC curves is
smaller) and at the same time coercive field values are few
thousands of Oe smaller than the cylinder-type shape values (circle
symbols Figure \ref{Fig_2}).

It is worth noticing at this point that even if we could not
extrapolate absolute magnetization values from our measurements for
the reason detailed above, we analyzed the
$\left(\frac{M_{r}}{M_{S}}\right)$ ratio. We defined as ususal
remanence as $M_{r}=M\left(H=0kOe\right)$ and, as the hysteresis
cycles have shown fully saturation around 30kOe, we defined the
saturation magnetization as $M_{s}=M\left(H=30kOe\right)$ (see inset
in Figure\ref{Fig_2}(a) in which only the half-positive part of the
hysteresis cycle is shown, the half-negative being symmetrical). We
found typical value of $\left(\frac{M_{r}}{M_{S}}\right)$
ratio$\approx0.5$ for all ZFC hysteresis loops. This latter value is
expected for a standard Stoner-Wohlfarth model in the case of
random-oriented nanowires \citet{Maurer_APL2007}. In the case of FC
hysteresis loops, the $\left(\frac{M_{r}}{M_{S}}\right)$ ratio was
usually$\geq0.6$ and reached the value of $\approx0.7$ for the
cylinder type nanowires (circle symbols), confirming the easier
alignment for the nanowires with the highest coercivity (see inset
\ref{Fig_2} (a)). We remind here that in the case of non-perfect
ellipsoidal morphology the remanence values for magnetically uniform
elongated nanoparticles are found to be lower then 1and may be as
low as 0.7 in the case of perfectly aligned dumbbells.\citet{F. Ott
- JAP,F. Zighem_S. Mercone}. Thus this may explains the small values
we observed even for the easier alignment.

\begin{figure}
\includegraphics[bb=2bp 330bp 250bp 590bp,clip,width=8.5cm]{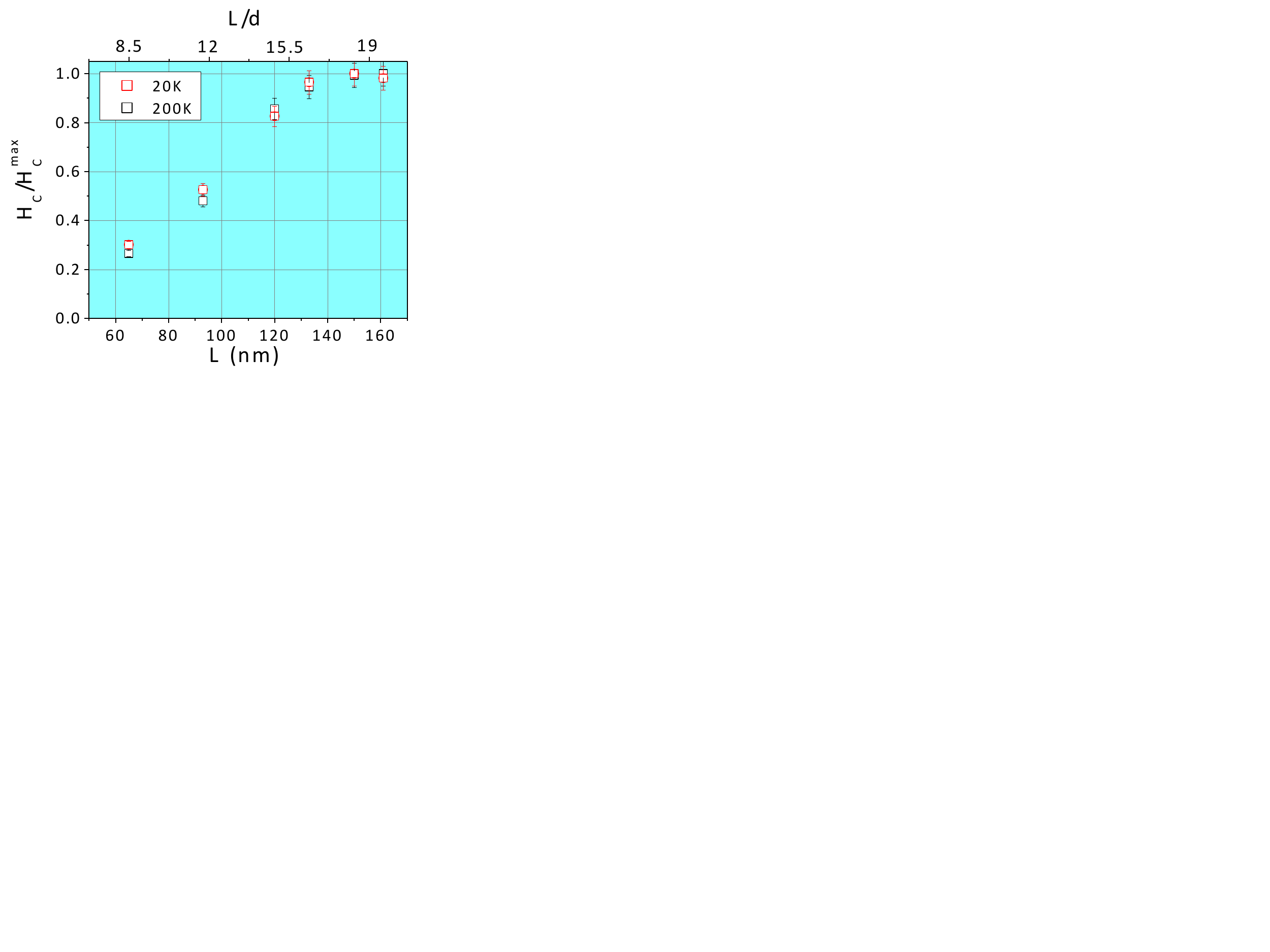}

\protect\protect\caption{Normalized coercive field at 20 K (red
circles) and 200 K (black squares) as function of the length for
diabolos nanowires having the same head morphology and diameter
(\textit{i.e.} $d$= 8 nm, $D$=12nm and $t$=8nm). Note that the
values have been normalized to the maximum coercivity observed for
the selected batches (\textit{i.e.} $H_{C}^{max}=5.8$ kOe). The top
scale reports the length over diameter ratio for the different
nanowires. }

\label{Fig_3}
\end{figure}

For obvious reasons of easy-axis configuration optmized along the
applied magnetic field, in order to study the morphology effect on
the coercive behavior observed in all our samples, we reported in
the bottom graph of the Figure \ref{Fig_2} the FC coercive field as
a function of temperature for various nanowire shapes. All of them
have ZFC coercive fields lower then the FC fields (not shown here),
in agreement with the behavior shown in the upper graph and
discussed previously. Micromagnetic simulations performed on
one-dimensional nano-objects of different shapes showed that the
magnetization reversal is strongly dependent of the length over
diameter ratio $\left(\frac{L}{d}\right)$\citet{F. Ott - JAP,F.
Zighem_S. Mercone}. From these previous works, the coercive values
in one-dimensional systems are expected to increase with the
increasing ratio for a very large range of length and diameter size.
This is confirmed by our measurements in the whole range of
temperature. This result is strengthened by the fact that the
expected behavior is observed for all head-morphology types. This
suggests that the ratio $\left(\frac{L}{d}\right)$ is the driving
parameter for the coercive field optimization, no matter the head
morphology. Thus, so far, the experimental data follow quite well
the standard model. In Ref. \citet{F. Zighem_S. Mercone},
micromagnetic simulations have been employed to demonstrate that
there is a critical value of the length over diameter ratio above
which the coercive field saturates. This result has been validated
by simulations of the reversal mechanism of various shapes
(\textit{i.e.} ellipsoid, cylinder, diabolo and dumbbell) and they
yielded a critical ratio $\frac{L}{d}=10$. In order to compare our
experimental results to this micromagnetic observation, we selected
among all the different batches, the samples showing the same head
morphology and the same diameter and we inspected their coercive
behavior as a function of the length. Figure \ref{Fig_3} presents
the normalized coercivity as function of the length for diabolo
nanowires having the same $d$= 8 nm, the same head-width $D$=12 nm
and the same head-thickness $t=8$ nm (equality are considered within
the statistical error of around $\pm1$ nm). The coercive field
values have been analyzed at low (20K) and high (200K) temperature.
These values clearly support the previous micromagnetic simulations
\citet{F. Zighem_S. Mercone} showing that the maximum coercive value
is constant for the longest nanowires and start to decrease for
$\frac{L}{d}\leq15$ (10\% loss is observed in Figure \ref{Fig_3} at
$\frac{L}{d}=15$ ).

\section{Head morphology effects and correction factor}

\begin{figure}
\includegraphics[bb=20bp 350bp 320bp 595bp,clip,width=8.5cm]{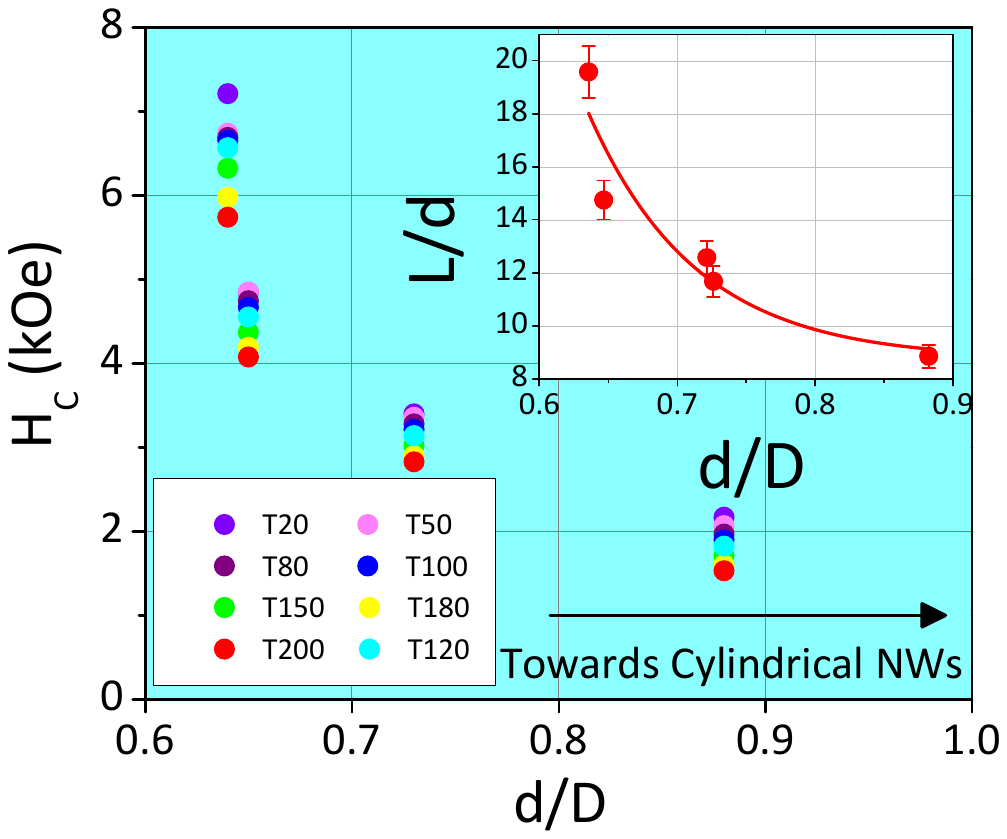}

\protect\protect\caption{Coercive field ($H_{C}$) variations (at
different temperature, see symbols for correspondences) as function
of the head-morphology ratio ($\frac{d}{D}$) showing the paradoxical
fact that nanowires with a diabolo shape are better nanomagnets then
the cylinder one. The insert corresponds to the one-dimensional
shape parameter $\frac{L}{d}$ as function of head-morphology
$\frac{d}{D}$ one.}

\label{Fig_4}
\end{figure}

In order to explore the morphology effect on the reversal mechanism
in nanomagnets, we decided to analyze the temperature dependence of
$H_{C}$ as function of the head morphology ratio $\frac{d}{D}$
(\textit{i.e.} the width of the conic head base ($D$ ) over the
diameter taken in the middle of the nanowire ($d$)). We performed
this analysis for nanowires with shapes varying from the
diabolo-type to the cylinder-type. Figure \ref{Fig_4} presents the
measured coercive fields as function of $\frac{d}{D}$ ratio. As
expected, at a fixed $\frac{d}{D}$ ratio, $H_{C}$ increases as
temperature is decreased for each head morphology (see colored
symbols in Figure \ref{Fig_4}). However, an unexpected
$H_{C}$-decrease of several kOe is observed approaching the
cylindrical shape while the diabolo-type nanowires (\textit{i.e.}
lower $\frac{d}{D}$ ratio) show higher coercivity values. It is
worth mentioning that micromagnetic simulations have already shown
that the cylindrical shape is a better geometry than the diabolo one
in order to maximize the coercive field \citet{F. Ott - JAP}. This
is not the case in our experimental observations in which nanowires
with more pronounced conic-head (\textit{i.e.} diabolo type) present
the higher $H_{C}$. At this point, we decided to test the
correlation between the $\frac{L}{d}$ ratio and the $\frac{d}{D}$
ratio of the nanowires. $\frac{L}{d}$ is reported as a function of
$\frac{d}{D}$ in the inset of Figure \ref{Fig_4}. Due to the
elaboration parameters explored during the nanowire growth, this two
ratios show a tight linked behavior. The shape ratio decreases with
the increasing head morphology parameter favoring the growth of a
pronounced head in longer nanowires. Looking at the results in the
inset of Figure\ref{Fig_4}, we thus expect from the nanowires having
the smaller $\frac{d}{D}(\lesssim0.65)$ ratio to present the higher
$H_{C}$ as these latter have also the higher $\frac{L}{d}$ ratio
value. Respectively, we expect from the nanowires having the highest
$\frac{d}{D}(\lesssim0.9)$ ratio to present the smaller $H_{C}$ as
these latter have also the lower $\frac{L}{d}$ ratio value. In the
main figure \ref{Fig_4}, this is exactly the coercivity behavior
observed. From these observations, we can conclude that the head
morphology ($\frac{d}{D}$) effect on the reversal mechanism, comes
directly from the shape ($\frac{L}{d}$) effect reported in the
previous paragraph. Due to specific growing conditions, the aspect
ratio $\frac{L}{d}$ and head morphology parameter $\frac{d}{D}$ are
not independent parameters and do not allow us to explore them
separately; this could lead to a paradoxical interpretation where a
diabolo-shaped nanowire is a better nanomagnet than a
cylinder-shaped one.

\begin{figure}
\includegraphics[bb=20bp 130bp 350bp 595bp,clip,width=8.5cm]{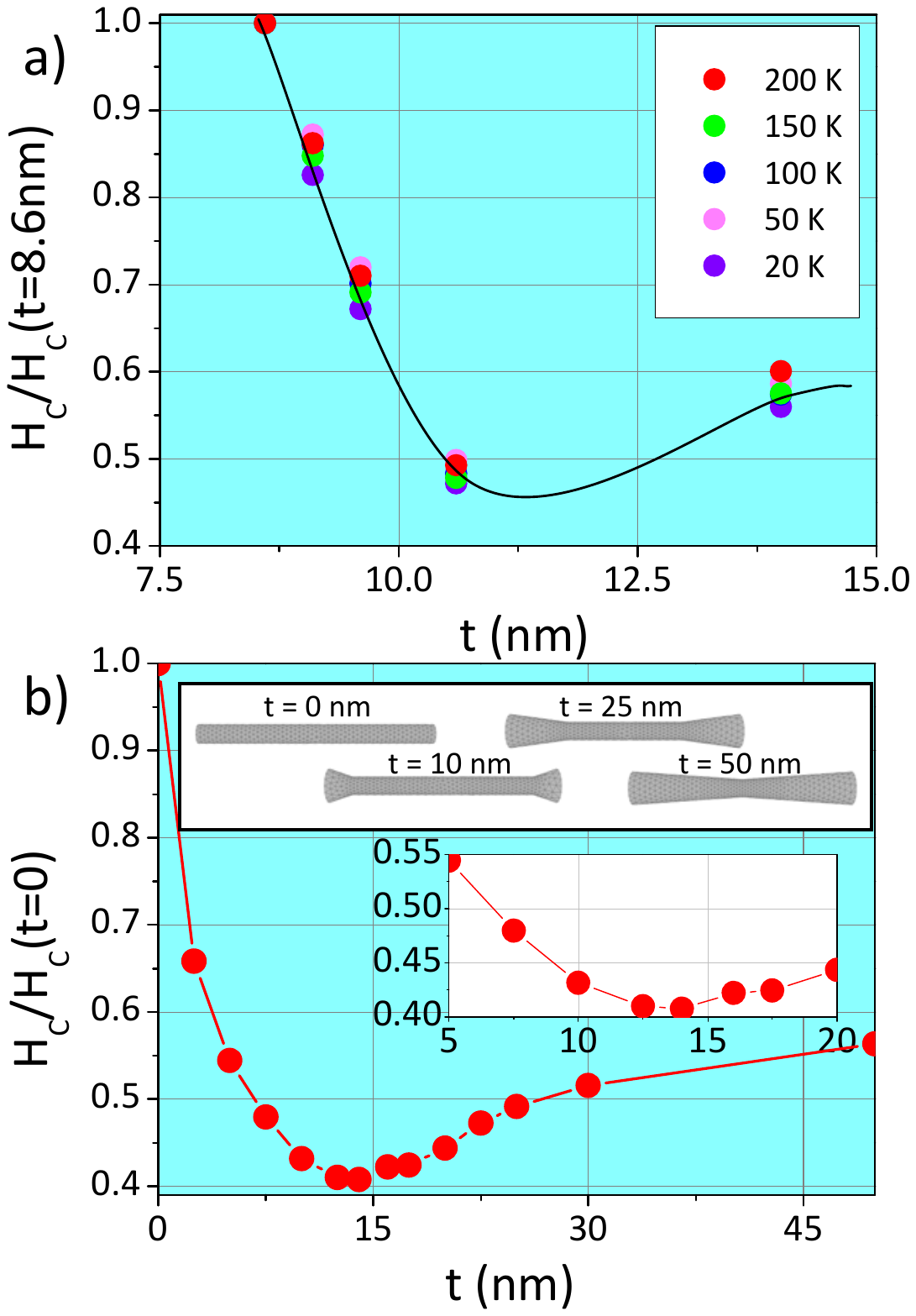}

\protect\protect\protect\caption{a) Normalized experimental coercive
field values for different temperatures versus head thickness
($\frac{L}{d}$ > 15); b) Normalized Coercive field values versus
head thickness for simulated nanowires with an aspect ratio $L/d=20$
(i.e. $L=200$ nm and $d=10$ nm) and a head width $D=20$ nm; meshes
of the nanowires for 4 different thickness ($t=0,10,25,50$ nm) are
shown in the upper part of the figure; the inset is a zoom around
the minimum of the curve.}

\label{Fig_5}
\end{figure}

In order to further explore the head morphology effects, we found it
opportune to analyze independently the effects of the head thickness
$t$. Thus the $H_{C}$-dependence with $t(nm)$ has been studied.
Figure \ref{Fig_5} a) presents the normalized $H_{C}$ variation as
function of $t$. The normalization of $H_{C}$ has been done by using
the coercive field of the smaller available thickness $t$
($H_{C}\left(t=8.6nm\right)$) at each temperature. This figure
clearly shows that the thicker is the conic head (\textit{i.e.}
higher $t$), the higher is the loss in coercivity. Surprisingly
enough, after reaching a minimum (around $t=10.5nm$) a rise in the
coercivity is then observed with increasing $t$.

In order to verify that the driving parameter of the unusual
behavior observed in Figure \ref{Fig_5} a) is the thickness $t$ of
the nanowire head, we performed micromagnetic simulation for
nanowires having the same average shape and morphology as the
measured ones ($L=150$ nm, $D=20$ nm, $d=10$ nm) and with a
thickness $t$ ranging from 0 to 50 nm. The simulations were
performed using the NMAG package. Details on procedure and
parameters used for these simulations can be found in elsewhere
\citet{F. Zighem_S. Mercone}. The magnetic parameters of bulk
Co80Ni20 were used during the simulations, namely: magnetization
saturation $M_{S}=1.2*10^{3}emu.cm^{-3}$and exchange stiffness
$A=1.2*10^{6}erg.cm^{-1}$. No magnetocrystalline anisotropy was
included to avoid mixing different sources of coercivity. In Figure
\ref{Fig_5} b), the simulated normalized
$H_{C}\left(t\left(nm\right)\right)$ curve qualitatively reproduced
the experimental results which proves that the unusual magnetic
behavior is driven by the $t$-variation. Previous works\citet{L.
Sun,F. Ott - JAP,F. Zighem_S. Mercone}demonstrated that for diabolo
and dumbbell shaped nanowires, the increasing width of the head
should yield a continuous decrease of the coercivity. Thus the
$H_{C}\left(D\left(nm\right)\right)$ theoretical behavior is
expected to monotonically decrease without any unexpected minimum.
From Figure 5b), we can intuitively understand that the rise of
$H_{C}$ at high $t$ is due to the tendency of the nanowires shape to
go towards a cylinder having a bigger diameter in the middle
$d=D=20$ nm (see schematics in Figure \ref{Fig_5} b)). This latter
has a lower coercive value (\textit{i.e.} $\sim0.6$ times the one of
the cylinder having $d=10$ nm) as reported in previous
works\citet{J. Zhang,H. Zeng,D. Ung,R.P. Cowburn,A. Gaul,F. Ott -
JAP,F. Zighem_S. Mercone}. These results are, to our knowledge, the
first experimental proof of the existence of a morphology factor
that corrects the magnetization reversal mechanism mainly driven by
the shape aspect ratio. The correction factor $t$ is independent
from the aspect ratio $\frac{L}{d}$ and drives the non-trivial
behavior observed in Figure \ref{Fig_5}.

\section{Conclusions }

In conclusion we reported the morphology dependence of the
magnetization reversal mechanism in weak interacting
(non-agglomerated) Co$_{80}$Ni$_{20}$ pure hcp nanowires.
Non-agglomerated nanowires with a controlled shape going from a
quasi perfect cylinder to a diabolo type have been studied
($D=\left(8-24\right)$ nm, $t=\left(7-14\right)$ nm,
$d=\left(7-12\right)$ nm). The coercive field versus the shape ratio
$(\frac{L}{d})$ followed a standard expected behavior simulated in
previous micromagnetic calculations: $H_{C}\left(\frac{L}{d}\right)$
increases with the increasing ratio. Our results confirm the
predominant effect of the shape ratio on the magnetization reversal
mechanism inside nanomagnets. Interestingly, an unexpected
paradoxical behavior was observed as function of the head morphology
($\frac{d}{D}$). $H_{C}$ decreases with increasing $\frac{d}{D}$,
leading to a picture in which a diabolo shaped nanowire is a better
nanomagnet than a perfect cylinder. This effect can be interpreted
by the strong correlation between the head morphology ratio
($\frac{d}{D}$) and the shape aspect one $(\frac{L}{d})$. Further
experimental analysis and micromagnetic simulations are in excellent
agreement showing a peculiar non-monotonous variation of the
coercivity field $H_{c}$ with the head thickness parameter $t$. In
conclusion, our results show clearly the existence of two
independent morphology parameters driving the magnetization reversal
mechanism in nanomagnets. This makes it stringent, for nanomagnet
application of cobalt based nanowires, to optimize both the dominant
shape aspect ratio parameter$(\frac{L}{d})$ and the non-trivial head
thickness independent one ($t$).
\begin{acknowledgments}
The magnetic measurements at ESPCI have been supported through
grants from Region Ile-de-France. ANR (AgenceNationale de la
Recherche) and CGI (Commissariat Général à l'Investissement) are
gratefully acknowledged for their financial support of this work
through Labex SEAM (Science Engineering for Advanced Materials and
devices) ANR 11 LABX 086, ANR 11 IDEX 05 02.
\end{acknowledgments}

\end{document}